# Si-based GeSn lasers with wavelength coverage of 2 to 3 μm and operating temperatures up to 180 K

Joe Margetis[1*], Sattar Al-Kabi[2*], Wei Du[3,4], Wei Dou[2], Yiyin Zhou[2,5], Thach Pham[2,5], Perry Grant[2,5], Seyed Ghetmiri[2], Aboozar Mosleh[2], Baohua Li[5], Jifeng Liu[6], Greg Sun[7], Richard Soref[7], John Tolle[1], Mansour Mortazavi[3], Shui-Qing Yu[2†]

[1]ASM, 3440 East University Drive, Phoenix, Arizona 85034, USA

[2]Department of Electrical Engineering, University of Arkansas, Fayetteville, Arkansas 72701 USA

[3]Department of chemistry and Physics, University of Arkansas at Pine Bluff, Pine Bluff, Arkansas 71601 USA

[4]Department of Electrical Engineering, Wilkes University, Wilkes-Barre, Pennsylvania 18766 USA

[5]Arktonics, LLC. 1339 South Pinnacle Drive, Fayetteville, Arkansas 72701 USA

[6]Thayer School of Engineering, Dartmouth College, Hanover, New Hampshire 03755 USA

[7]Department of Engineering, University of Massachusetts Boston, Boston, Massachusetts 02125 USA.

*These authors contributed equally to this work.
†Corresponding author.

**Abstract**

A Si-based monolithic laser is highly desirable for full integration of Si-photonics. Lasing from direct bandgap group-IV GeSn alloy has opened a completely new venue from the traditional III-V integration approach. We demonstrated optically pumped GeSn lasers on Si with broad wavelength coverage from 2 to 3 μm. The GeSn alloys were grown using newly developed approaches with an industry standard chemical vapor deposition reactor and low-cost commercially available precursors. The achieved maximum Sn composition of 17.5% exceeded the generally acknowledged Sn incorporation limits for using similar deposition chemistries. The highest lasing temperature was measured as 180 K with the active layer thickness as thin as 260 nm. The unprecedented lasing performance is mainly due to the unique growth approaches, which offer high-quality epitaxial materials. The results reported in this work show a major advance towards Si-based mid-infrared laser sources for integrated photonics.

Si-based electronics industry has driven the digital revolution for an unprecedented success. As a result, there has been tremendous effort to broaden the reach of Si technology to build integrated photonics[1-3]. Although great success has been made on Si-based waveguides[4], modulators[5], and photodetectors[6,7], a monolithic integrated light source on Si with high efficiency and reliability remains missing and is seen as the most challenging task to form a complete set of Si photonic components. Currently, Si photonics utilizes direct bandgap III-V lasers as the light source through different integration approaches such as wafer-bonding or direct-growth, which has seen significant progress in the last decade[8-11]. From the other side, a laser made from direct bandgap group-IV materials offers its own unique advantages particularly the material integration compatibility with the complementary metal–oxide–semiconductor (CMOS) process. However, group-IV materials such as Si, Ge, and SiGe alloys have been excluded from being an efficient light source due to their indirect bandgap nature. Although the optically pumped Er doped Si laser[12] and the Si Raman laser[13] have been reported, they do not rely on bandgap emission and cannot be operated under direct electrical pumping. Recently developed Ge lasers[14-16] employed strain-engineering and heavily n-type doping to compensate for its indirect bandgap, yet the high threshold and fabrication difficulties remain unresolved thus far.

It has been theoretically predicted that the group-IV alloy GeSn could achieve a direct bandgap by incorporating more than 6-10% Sn into Ge[17-19]. In order to overcome the limit of low solid solubility of Sn in Ge (<1%), low temperature growth techniques under non-equilibrium conditions have been successfully developed[20,21], leading to the first experimental demonstration of direct bandgap GeSn alloy[22] and the optically pumped GeSn interband lasers[23-25]. The recent engagement of mainstream industrial chemical vapor deposition (CVD) reactors for the development of GeSn growth techniques has enabled those significant results, which implies that the growth method is manufacturable and can be transferred to the foundry/fab[3]. In this paper, we demonstrate the first set of optically pumped GeSn edge-emitting lasers that covers an unprecedented broad wavelength range from 2 to 3 μm with lower lasing threshold and higher operation temperature than all previous reports. This superior laser performance is attributed to the unique epitaxial growth approaches that were developed based on newly discovered growth dynamics. Contrary to the common belief that growing GeSn with high Sn composition is mainly limited by the chemistry of the deposition process, we found that the Sn incorporation is actually limited by the compressive strain. Strain-induced Sn segregation results in continuous ejection of Sn solute atoms which prevents increasing the Sn concentration through the adjustment of the $SnCl_4$ partial pressure alone. By relieving the strain constraint, a much higher Sn composition with high crystallinity was obtained.

It is generally acknowledged that a higher Sn composition (more than 10%) is preferred to obtain high performance GeSn lasers as higher Sn composition enables more bandgap directness and enhances a more favorable direct radiative recombination. For GeSn growth using $SnCl_4$ and different Ge-hydrides as precursors, growth dynamics studies revealed that lower growth temperature is required for higher Sn incorporation[26-30]. Currently, the Sn compositions up to 12% were obtained by using $GeH_4$ as precursor[26,27] while the Sn compositions up to 15% were achieved by using $Ge_2H_6$ or $Ge_3H_8$ as precursor[28-30]. Using $GeH_4$ is preferred from an industrial manufacturing perspective due to its much lower cost. However, use of the $Ge_2H_6$ and $Ge_3H_8$ remains attractive due to the fact that the high order hydrides decompose more readily at lower temperatures, which facilitates the growth of higher Sn composition GeSn. Although significant progress has been made, maximum Sn compositions using industrial CVD methods seemed to have plateaued in the ~15% Sn range regardless of precursor choice or growth recipe specifics. Such a limit has been mainly attributed to the chemical reaction balance dominating the growth process and the availability, or lack thereof, of disassociated Ge-hydrides and Sn-chlorides as the temperature is continually decreased to counteract Sn out-diffusion/segregation[28-30].

In our recent work, we have observed a clear spontaneous-relaxation-enhanced (SRE) Sn incorporation process. When GeSn is grown on a Ge buffer using a nominal 9% $GeH_4$ based recipe, the Sn incorporation starts from 9% and the material gradually relaxes and then the subsequent GeSn layer Sn composition changes to 12%. The growth normally results in a distinct two-layer structure with the first layer being defective and gradually relaxed and the second layer being low-defect density and almost completely relaxed. The fact that the growth recipe is maintained the same for the entire SRE growth process strongly suggests that the *compressive strain* rather than the chemical reaction is the dominant limiting factor of Sn incorporation[24]. This discovery inspires us to adopt two approaches as the new growth strategy to obtain high Sn compositions: i) the SRE approach and ii) the GeSn virtual substrate (VS) approach, which lead to the final Sn composition up to 15.9% and 17.5%, respectively. The GeSn VS approach utilizes the GeSn layers obtained through the SRE approach as the buffer to grow higher Sn composition films with an optimized recipe. The relaxed template allows for higher $SnCl_4$ partial pressures to be used which when directly applied to growth on a Ge buffer, would cause strain-induced Sn segregation and precipitation.

In this work, two sets of GeSn samples were grown using an industry-standard ASM Epsilon® 2000 PLUS reduced pressure chemical vapor deposition (RPCVD) reactor[27]. As listed in Table 1: samples A to E were grown via the SRE approach (two-layer GeSn structure) while samples F and G were grown via the GeSn VS approach (two layer GeSn structure plus an additional third GeSn layer). Note that sample B was capped with a 10 nm Ge passivation layer which lately showed negligible effects for the overall device performance. A standard Ge buffer layer was grown on Si *in-situ* prior to the GeSn deposition. After the growth, the GeSn layer thickness and quality in terms of threading dislocation density (TDD) were analyzed using transmission electron microscopy (TEM) and etch pit density techniques. The Secondary Ion Mass Spectrometry, X-ray diffraction (XRD) 2θ-ω scan and reciprocal space mapping (RSM) were used to determine the Sn compositions and strain after cross check, based on which the electronic bandgap structures at room temperature were calculated. The detailed material characterization results are also summarized in Table 1.

**Table 1** Summary of sample material and lasing characterization results

| # | GeSn 1st layer* | | GeSn 2nd layer | | GeSn 3rd layer** | | $T_0$ (K) | Lasing wavelength @ 77 K (nm) | Threshold @ 77 K (kW/cm²) |
|---|---|---|---|---|---|---|---|---|---|
| | Sn% | Thickness (nm) | Sn% | Thickness (nm) | Sn% | Thickness (nm) | | | |
| A | 5.6% | 210 | 7.3% | 680 | | | N. A. | 2070 | 300 |
| B | 8.3% | 280 | 9.9% | 850 | | | 76 | 2400 | 117 |
| C | 9.4% | 180 | 11.4% | 660 | | | 87 | 2461 | 160 |
| D | 10.5% | 250 | 14.4% | 670 | | | 73 | 2627 | 138 |
| E | 11.6% | 210 | 15.9% | 450 | | | N. A. | 2660 | 267 |
| F | 9.8% | 160 | 12.7% | 680 | 16.6% | 290 | 84 | 2767 | 150 |
| G | 11.9% | 310 | 15.5% | 550 | 17.5% | 260 | 73 | 2827 | 171 |

*The "GeSn 1st layer" is directly in contact with the Ge buffer and the Sn composition is the initial nominal value. It gradually relaxes to a composition close to the composition in the "GeSn 2nd layer;
**The GeSn 3rd layer was observed from samples F and G.

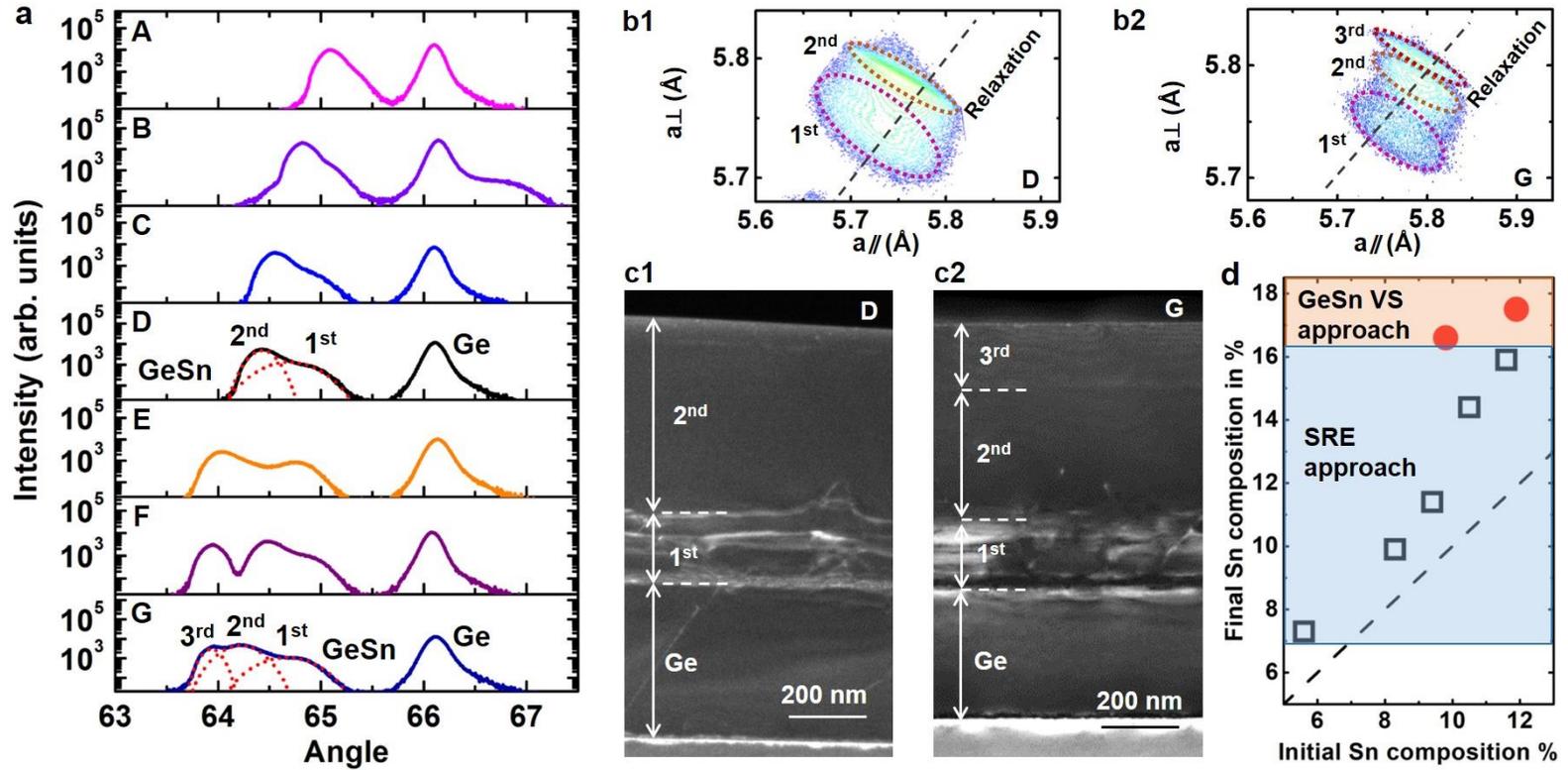

**Figure 1.** Material characterization of samples. (a) The 2θ-ω scan showing the gradually shifted GeSn peaks with increased Sn%. The fitted peaks indicate the existence of two and three GeSn layers for samples D and G, respectively. (b1) and (b2) RSM contour plots of sample D and G indicating the relaxed multiple GeSn layers. (c1) and (c2) TEM images of the samples D and G showing high quality GeSn top layers and the formation of TD loop in the 1st GeSn layers. (d) Summary of the material growth. The relaxed GeSn results in more incorporated Sn during the following growth. The Sn composition of 17.5% was achieved with stepped GeSn buffer growth.

Figure 1a shows the XRD 2θ-ω scan with all curves aligned with the Si peak. The peaks at 66.1° are attributed to the Ge buffer while the peaks between 65.5°-63.5° belong to GeSn layers. As the Sn composition increases, the GeSn peak shifts towards a lower angle. The asymmetric GeSn peak for each sample indicates the existence of multiple layers corresponding to different Sn compositions in the GeSn film. Based on a multi-peak fitting process (Supplementary

Section 1), for the samples A to E, two distinct GeSn peaks can be obtained (See sample D as an example); while for the samples F and G, three peaks were clearly resolved (fitting curves for sample G are shown). The multi-GeSn-layer characteristic was also observed in RSM and TEM images. Figures 1-b1 and b2 shows the RSM contour plots of samples D and G indicating the two-layer and three-layer characteristics, respectively. The dashed ellipse annotates each GeSn layer, which features an almost complete strain relaxation. The high resolution TEM images of samples D and G are shown in Fig. 1-c1 and c2. For the Ge buffer layer in each sample, the majority of defects were localized in the Ge/Si interface so that the high-quality Ge buffer was obtained. For the GeSn alloy, each layer can be clearly observed. Note that for each sample, the 1st GeSn layer over the Ge buffer is defective due to the high TDD, which is attributed to the lattice mismatch between the Ge buffer and the 1st GeSn layer. However, the formation of threading dislocation loops in the 1st GeSn layer prevents the defects from being propagated to the 2nd GeSn layer of sample D and to the 2nd and 3rd GeSn layers of sample G, resulting in low-defect density GeSn top layers. The mechanism of formation of dislocation loops during GeSn growth could be attributed to the special crystallographic plane and the external shear stress[31]. By using TEM and etch pit density measurements, a TDD of $10^6$ cm$^{-2}$ was obtained for the GeSn top layer while the TDD of Ge buffer layer (in a separate control sample with the same Ge thickness) was measured as $10^7$ cm$^{-2}$.

Figure 1d summarizes the GeSn growth results that were attained using the newly developed approaches. For instance in the SRE approach, using a nominal 10.5% recipe for a total 920-nm-thick growth (sample D), the Sn compositions were measured as 10.5% and 14.4% in the 1st and 2nd layers (250 and 670 nm thick), respectively. Using this approach, the 1.6%~4.3% increase in Sn composition for the 2nd layer compared to that in the 1st layer was achieved for all samples. For the growth using the GeSn VS approach, for example sample G, a relaxed GeSn buffer with a final Sn composition of 15.5% was first obtained and then followed by a top GeSn layer with a much higher Sn composition of 17.5%, breaking the record of Sn incorporation not only for GeH$_4$-based CVD recipe (12%)[27], but also Ge$_2$H$_6$-based recipe (15%)[30].

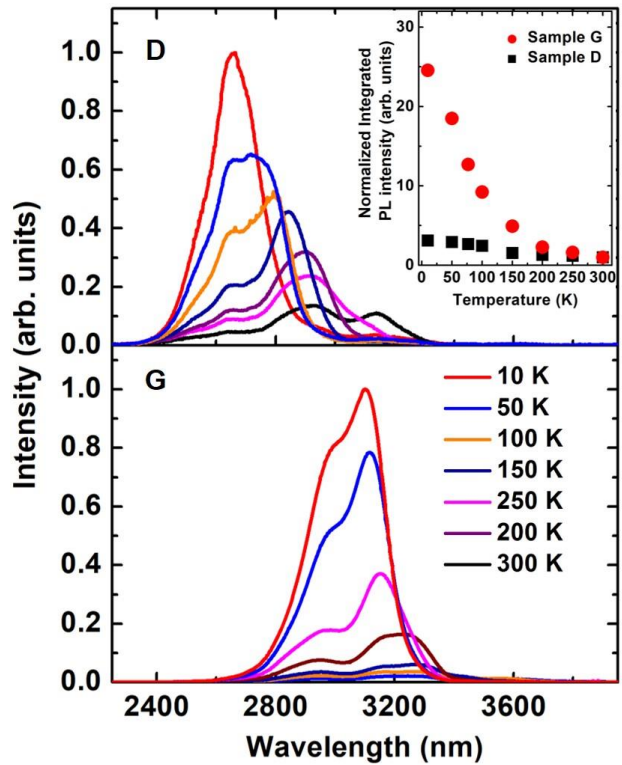

**Figure 2.** Temperature-dependent PL spectra of samples D and G. The dramatically increased PL intensity at lower temperature indicates the direct bandgap material nature. Inset: Normalized integrated PL intensity. The additional valley feature appearing at ~3.0 μm in all spectra is due to the water absorption.

The temperature-dependent photoluminescence (PL) measurements were performed to confirm the bandgap directness of the GeSn layers. For each sample, as the temperature decreases from 300 K to 10 K, the PL intensity dramatically increases, showing the clear signature of the direct bandgap similar to that of III-V materials such as GaAs and InP. The typical temperature dependent PL spectra for samples D and G are shown in Fig. 2. The observed PL peak blue-shift at lower temperature is expected due to the increase of the bandgap. From 300 to 10 K, sample D features a 4-times increase of integrated PL intensity while sample G features a 25-times increase of integrated PL intensity, as shown in Fig. 2 inset. The distinct difference is attributed to enhanced direct bandgap radiative recombination due to a higher Sn composition in sample G. It is worth noting that since the penetration depth of the 532-nm excitation laser beam is less than 100 nm, the optical transitions including light absorption and emission can occur *only in the top GeSn layer* (i.e. the 2$^{nd}$ layer of sample D and the 3$^{rd}$ layer of sample G). To further investigate the optical transition property, a 1060-nm laser with penetration depth of over 1 μm was used. Very similar PL spectra were obtained, indicating that the PL emissions are mainly from the band-to-band recombination in the top GeSn layer. This can be interpreted as the following: since the top GeSn layer features a narrower bandgap compared to the GeSn layer(s) underneath due to higher Sn composition and a type-I band alignment, the carriers are confined in the top GeSn layer to recombine. The carrier confinement effect was further confirmed by the band diagram calculation results (Supplementary Section 2). These results also point to a promising path to achieve type-I quantum well lasers in the GeSn system.

    Ridge-waveguide-based edge-emitting lasers were fabricated by standard lithography and etching processes. A low temperature wet chemical etching process was developed in this work. By using a mixture of HCl: $H_2O_2$: $H_2O$=1:1:10 at 0 °C, an average etching rate of ~20 nm/min was obtained, which is almost a constant regardless of the Sn composition. Due to the lateral

etching, the width of the top and bottom of the ridge waveguide were measured as 2 and 5 μm, respectively. The etching depth was selected as 800 nm to provide a sufficient mode confinement. After etching, the samples were lapped down to ~70 μm thickness and then cleaved to different cavity length to finish the devices. Figures 3a and 3b show the schematic waveguide structure and cross-sectional view of the scanning electron microscope (SEM) image. Atomic-force microscopy (AFM) characterizations have been performed on as-grown samples and post-etch samples (Supplementary Section 4) with values ranged from 3.75 to 18.00 nm, and 6.27 to 13.70 nm, respectively. The surface roughness of post-etch samples is estimated to only result in moderate scattering loss and therefore the laser performance would not be deteriorated. Figure 3a inset indicates an overlap of the fundamental TE mode with the GeSn layer for a confinement factor of 85.2% (Supplementary Section 5).

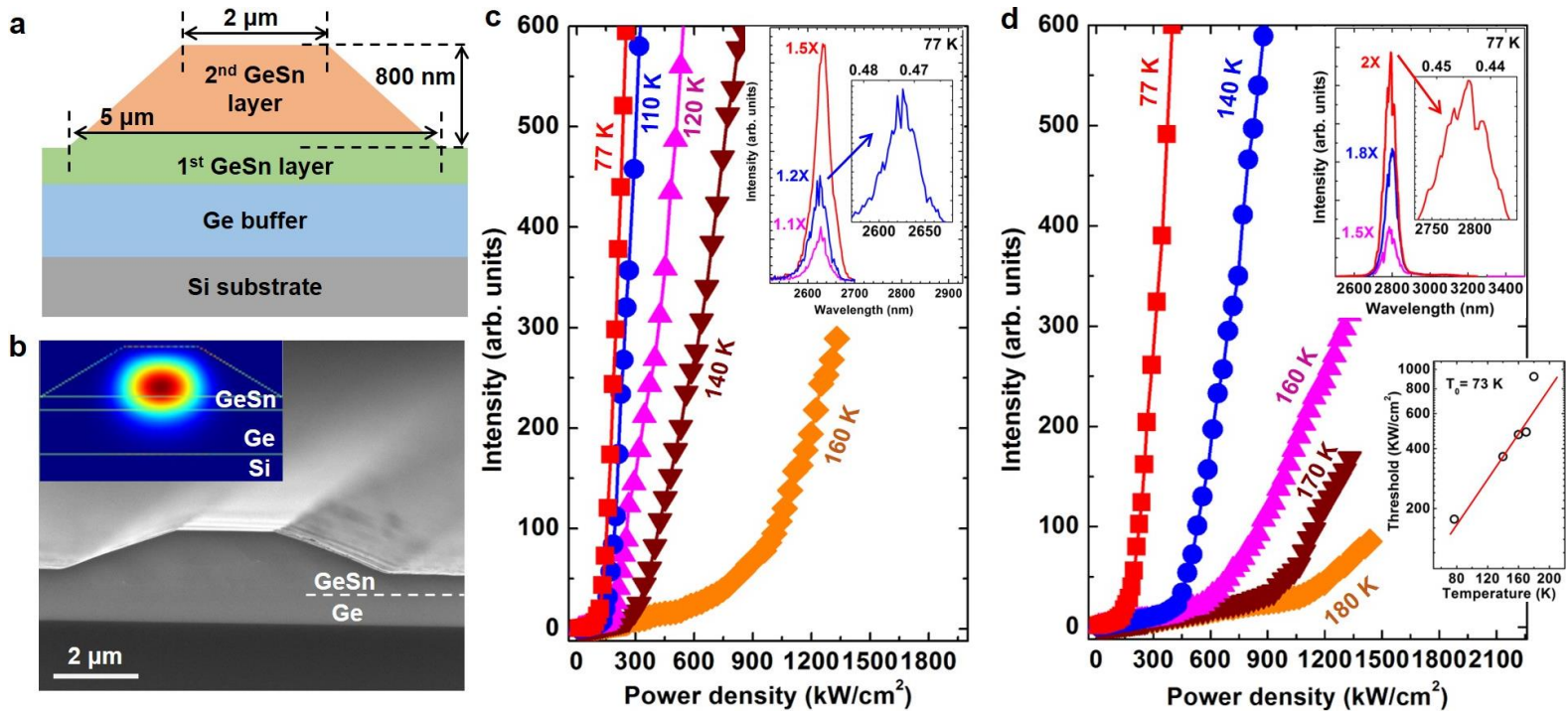

**Figure 3.** GeSn optically pumped laser using samples D and G. (a) Schematic of layer structure (not to scale). (b) Cross-

sectional view of SEM shows device mirror-like facet. Inset: fundamental TE mode profile. (c) and (d) Light output power versus pump laser power at various temperatures. Inset: Lasing spectra at 77 K indicating the multi-mode behavior (upper x-axis shows energy in eV); characteristic temperature ($T_0$) of sample G.

The optical pumping characterization was performed using a pulsed 1060 nm laser with 45 kHz repetition rate and 6 ns pulse width. The laser-output versus pumping-laser-input (L-L) curves for each sample were measured from 77 K to each individual maximum operating temperature. At 77 K, the lasing thresholds for all samples were obtained ranging from 117 to 300 kW/cm$^2$, as listed in Table 1. The relatively high lasing-onset excitation of 300 kW/cm$^2$ of sample A is mainly due to the lower Sn composition relative to other samples, which results in less favorable populating of electrons in the direct $\Gamma$ valley under the same pumping power density. The lasing threshold of 267 kW/cm$^2$ for sample E might be due to the slightly degraded material quality as it features the highest Sn composition (15.9%) among those samples grown via the SRE approach. The relative lower quality of sample E was confirmed by XRD characterizations, which indicate that sample E features larger peak line-width of 2θ-ω scan and broadened contour plot of RSM compared to other samples. The thresholds of the remaining five samples ranging from 117 to 171 kW/cm$^2$ at 77 K are lower than that of our previously demonstrated GeSn laser[24] (~200 kW/cm$^2$).

Figure 3c and 3d show the L-L curves for sample D (550 μm-long device) and G (700 μm-long device). For each curve the threshold characteristic was clearly observed. The lasing wavelengths were observed at 2630 and 2845 nm at 77 K for samples D and G, respectively. The maximum lasing temperatures were measured as 160 and 180 K for samples D and G, with the corresponding thresholds of 795 and 920 kW/cm$^2$, respectively. Both maximum operating temperatures are higher than that of the reported GeSn micro-disk lasers[25] (130 K) and ridge waveguide lasers (110 K).[24] We attribute the superior laser performance to the high-quality materials that were grown via unique approaches, and to the significant reduction of the modal overlap with the misfit dislocations localized at the Ge/GeSn interface[32].

The investigation of the lasing mode characteristics was performed via lasing spectra measurement. Due to the relatively large area of the cavity facet, the lasing spectra exhibit a typical multimode lasing characteristic. The lasing spectra at 77 K for samples D and G are shown in Fig. 3c and 3d insets. At the 1.2-times lasing threshold excitation power density for sample D, the full width at half maximum (FWHM) of each resolved peak was estimated ranging from 6 to 11 nm (1.3 to 2.4 meV). On the other hand, at 2.0-times lasing threshold for sample G, the FWHM of each resolved peak was extracted ranging from 9 to 12 nm (1.9 to 2.6 meV). Compared with the FWHM of PL spectra at 10 K shown in Fig. 2, which are 114 and 122 nm (22.6 and 21.2 meV) for samples D and G, respectively, the dramatically reduced line-widths shown in Fig. 3c and 3d insets indicates obvious evidence of lasing. At a pumping power density slightly higher than lasing threshold, the multi-peaks revealing the lasing modes can be observed clearly. As the pumping power increases, the modes become more pronounced and most peaks grow, resulting in the overall lasing intensity increase.

The characteristic temperature for each sample was extracted by temperature-dependent lasing threshold except for samples A and E due to insufficient data points, as listed in Table 1. As an example, Figure 3d inset (bottom) presents the fitted characteristic temperature of 73 K between 77 and 180 K for sample G. For the samples studied in this work, their characteristic temperatures ranging from 73 to 87 K show slight fluctuation, which could benchmark the current phase of development of GeSn material for laser applications. In comparison, the characteristic temperatures of earlier developed InP and GaAs based optically pumped lasers were reported as 100 and 129 K, respectively[33,34].

The laser emission spectra were fully studied and are summarized in Fig. 4a. At 77 K, lasing spectra from each sample were observed. As temperature increases, except for samples A (lowest Sn composition) and E (slightly degraded material quality), the GeSn lasers fabricated from other samples lased up to 140 K. In particular, samples D and G lased at 160 K, and the maximum lasing temperature is 180 K which was observed from sample G. The main factors that determine the lasing performance are the material gain, the active layer thickness, the surface roughness of the device, and the dominant nonradiative recombination, which vary among the samples investigated in this work at different temperatures, resulting in each individual maximum lasing temperature. It is worth noting that the early theoretical prediction for the highest achievable lasing temperature for the GeSn double heterostructure (DHS) laser was 200 K[35], beyond which the Auger recombination could be dominant and stop the lasing process. With the current lasing temperature being close to the theoretical predication, it is essential in the future to switch from the bulk DHS to more advanced device structures such as utilizing multiple GeSn quantum wells for room temperature operation[35-37], as suggested in ref. 35 that with a possible active region design of 20 SiGeSn/GeSn QWs. As we have shown previously in the PL results, the type-I alignment between the GeSn layers of different compositions provides a promising path to the quantum well lasers.

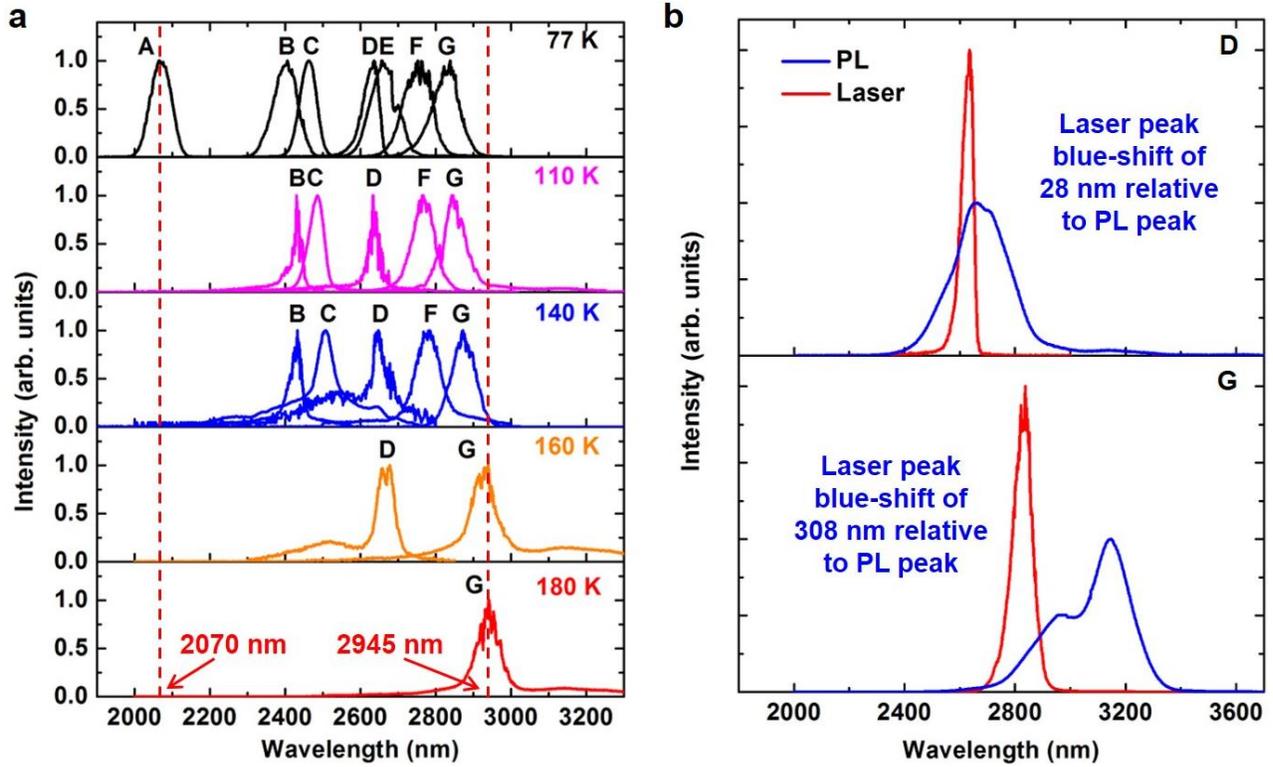

**Figure 4.** GeSn laser performance characterization. (a) Spectra of GeSn lasers fabricated from samples A to G at temperatures from 77 to 180 K. (b) Lasing spectra at 77 K of samples compared with those PL spectra. The lasing peak blue-shift is due to the typical band filling effect.

By increasing the final Sn compositions from 7.3% to 17.5%, a broad coverage of lasing wavelengths from 2 to 3 μm was achieved (2070 to 2945 nm). Note that the lasing wavelengths were determined not only by Sn composition but also strain status, operating temperature, active region thickness, and carrier population profile. To the best of our knowledge, this is the first demonstration of GeSn laser sets with such broad wavelength coverage. Moreover, the lasing at 2945 nm (Sample G at 180 K) is the longest emission wavelength reported so far from GeSn based lasers. In fact, there is no fundamental limit to extend lasing spectra towards even longer wavelengths via the unique material growth approach developed in this work to further increase Sn incorporation.

Further analysis on lasing behavior was conducted by the study of emission peaks comparison between PL and lasing, as plotted in Fig. 4b. For samples D and G, the lasing peaks showed a blue-shift of 28 and 308 nm relative to their PL peaks at 77 K, respectively. This blue-shift can be interpreted as the typical band filling effect, which is commonly observed from III-V lasers such as GaAs- and InP-based lasers with their relatively thinner gain layer. Sample G features more blue-shift relative to sample D. This is mainly due to the thinner GeSn top layer of sample G ($3^{rd}$ GeSn layer of 260 nm) compared to that of sample D ($2^{nd}$ GeSn layer of 670 nm). Therefore, for sample G the carriers would be confined in a small region near the top surface due to the nature of built-in type-I band alignment between the $3^{rd}$ and $2^{nd}$ GeSn layers, leading to a thinner active gain layer with higher injected carrier density, and consequently resulting in carriers populating at higher energy states and a more pronounced blue-shift at the intense pumping condition. A longer wavelength beyond 3 µm could be accessible if the third layer GeSn in sample G is increased to further increase the modal gain and reduce the band filling. Observing band filling for given GeSn thickness combined with knowledge of the band alignment for the current GeSn heterostructures provides a useful insight for the design of future quantum well based GeSn lasers by choosing the right well/barrier materials as well as the number of the wells.

In conclusion, we have demonstrated the GeSn optically pumped lasers covering the lasing wavelengths from 2 to 3 µm. The GeSn samples were grown using an industry standard CVD reactor with low-cost $SnCl_4$ and $GeH_4$ precursors. Newly discovered epitaxial growth dynamics indicate that the Sn incorporation is primarily limited by strain-induced Sn-segregation, with the deposition chemistry being secondary. By utilizing spontaneous-relaxation-enhanced Sn incorporation and GeSn virtual substrate approaches, a maximum Sn composition of 17.5% was achieved, which breaks the previous suspected 15% Sn limit even for using high order hydrides as precursors. Ridge waveguide-based edge-emitting F-P lasers were fabricated. Lasing from 2070 to 2945 nm was achieved with the maximum lasing temperature of 180 K. At 77 K, the lasing onset excitation of 117 kW/cm$^2$ was obtained. Significant lasing peak blue-shift relative to the PL peak was observed, reflecting the band filling effect, which is a typical characteristic of direct bandgap lasers such as those made from III-Vs. This work is an essential step towards obtaining high performance Si-based monolithic integrated mid-infrared laser sources.

METHODS

**Ge buffer growth.** A nominal 600 nm-thick Ge buffer layer was firstly grown on Si followed by an *in-situ* annealing to reduce the defects. The growth of Ge on Si proceeded via a Stranski–Krastanov mechanism in which an initial ~3 continuous monolayers were formed followed by the formation of islands. In this process the initial growth was conducted at low temperatures to extend the 2-D growth followed by a high temperature growth in which the bulk of the film was deposited. The higher temperature growth reduced the dislocations by promoting glide and subsequent annihilation of threading defects as well as providing increased growth rate. Further defect reduction can be accomplished by *in-situ* annealing of the layer. Utilizing this approach the relaxed Ge buffer layer with threading defect density of ~1×10$^7$ cm$^{-2}$ was obtained.

**Photoluminescence (PL) measurements.** The PL measurements were performed using a standard off-axis configuration with a lock-in technique (optically chopped at 377 Hz). A continuous wave (CW) laser emitting at 532 nm wavelength was used as an excitation source.

The laser beam was focused down to a 100 μm spot and the power was measured to be 100 mW. The PL emission was collected by a spectrometer and then sent to a PbS detector (cut-off at 3.0 μm, higher signal-to-noise ratio) or a InSb detector (cut-off at 5.0 μm, lower signal-to-noise ratio).

**Optical pumping measurements.** The optical pumping characterization was performed using a pulsed laser operating at 1060 nm with 45 kHz repetition rate and 6 ns pulse width. The laser beam was collimated to a narrow stripe (~20 µm width and 0.3 cm length) via a cylindrical lens to pump the GeSn waveguide structure. Since the spatial intensity profile of the laser beam features Gaussian distribution, the knife-edge technique was applied to determine the pumping power density. The device was first mounted on a Si chip carrier and then placed into a continuous flow cryostat for low temperature measurement. The emission from the facet was collected by a spectrometer and then sent to a PbS or InSb detector. The integrated emission intensity was measured by setting the grating at zero order.

AUTHOR INFORMATION

**Corresponding Author**

*E-mail: syu@uark.edu

**Author contributions**

S.Y., M.M., B.L., and W.D. proposed and guided the overall project. J.M., and J.T. planned and conducted the GeSn epitaxial growth. W.D., P.G., and A.M. performed material characterization. Y.Z., and T.P. fabricated the laser devices. S.G., and S.A. developed and conducted the optical and lasing measurements. W.D., Y.Z., J.L., G.S., and R.S. were involved with lasing evaluation, band diagram and mode calculations. All authors discussed the results and commented on the manuscript.

ASSOCIATED CONTENT

**Supporting information**

Supplementary information is available in the online version of the paper. Reprints and permissions information are available online at www.nature.com/reprints. Correspondence and requests for materials should be addressed to S.Y.

ACKNOWLEDGEMENTS

The authors acknowledge financial support from the Air Force Office of Scientific Research (Grant Nos. FA9550-14-1-0205, FA9550-14-1-0147, FA9550-16-C-0016, FA2386-14-1-4073), the National Science Foundation (Grant No. DMR-1149605) and Asian Office of Aerospace Research and Development (Grant No. FA2386-16-1-4069). We are thankful for the technical support from Institute for Nanoscience & Engineering, University of Arkansas, Dr. M. Benamara's assistance in TEM imaging and Dr. A. Kuchuk's assistance in XRD measurements.

REFERENCES

1. Soref, R. The past, present, and future of silicon photonics. *IEEE J. Sel. Top. Quantum Elect.* **12**, 1678–1687 (2006).

2. Soref, R. Mid-infrared photonics in silicon and germanium. *Nature Photon.* **4**, 495–497 (2010).

3. Soref, R., Buca, D., & Yu, S. -Q. Group IV Photonics: Driving Integrated Optoelectronics. *Opt. Photon. News* **27**, 32–39 (2016).

4. Soref, R., Emelett, S. J. & Buchwald, W. R. Silicon waveguided components for the long-wave infrared region, *J. Opt. A-Pure. Appl. Op.* **8**, 840–849 (2006).

5. Reed, G. T., Mashanovich, G., Gardes, F. & Thomson, D. Silicon optical modulators. *Nature Photon.* **4**, 518–526 (2010).

6. Michel, J., Liu, J. & Kimerling, L. C. High-performance Ge-on-Si photodetectors. *Nature Photon.* **4**, 527–534 (2010).

7. Pham, T. N. *et al*. Si-based $Ge_{0.9}Sn_{0.1}$ photodetector with peak responsivity of 2.85 A/W and longwave cutoff at 2.4 μm. *Electron. Lett.* **51**, 854–856 (2015).

8. Tanabe, K., Watanabe, K. & Arakawa, Y. III–V/Si hybrid photonic devices by direct fusion bonding. *Sci. Rep.* **2**, 349 (2012).

9. Wang, Z. et al. Room-temperature InP distributed feedback laser array directly grown on silicon. *Nature Photon.* **9**, 837–842 (2015).

10. Liu, H. *et al*. Long-wavelength InAs/GaAs quantum-dot laser diode monolithically grown on Ge substrate. *Nature Photon.* **5**, 416–419 (2011).

11. Zhou, Z. *et al*. On-chip light sources for silicon photonics. *Light Sci. Appl.* **4**, e358 (2015).

12. Kippenberg, T. J., Kalkman, J., Polman, A. & Vahala, K. J. Demonstration of an erbium-doped microdisk laser on a silicon chip. *Phys. Rev. A* **74**, 051802 (2006).

13. Rong, H. *et al*. A continuous-wave Raman silicon laser. *Nature* **433**, 292–294 (2005).

14. Liu, J. *et al*. Ge-on-Si laser operating at room temperature. *Opt. Lett.* **35**, 679–681 (2010).

15. Camacho-Aguilera, R. E. *et al*. An electrically pumped germanium laser. *Opt. Express* **20**, 11316–11320 (2012).

16. Koerner, R. *et al*. Electrically pumped lasing from Ge Fabry-Perot resonators on Si. Opt. Express **23**, 14815–14822 (2015).

17. Du, W. *et al*. Competition of optical transitions between direct and indirect bandgaps in $Ge_{1-x}Sn_x$. *Appl. Phys. Lett.* **105**, 051104 (2014).

18. Alberi, K. *et al*. Band anticrossing in highly mismatched $Sn_xGe_{1-x}$ semiconducting alloys. *Phys. Rev. B* **77**, 073202 (2008).

19. D'Costa, V. R. *et al*. Optical critical points of thin-film $Ge_{1-y}Sn_y$ alloys: A comparative $Ge_{1-y}Sn_y/Ge_{1-x}Si_x$ study. *Phys. Rev. B* **73**, 125207 (2006).

20. Oehme, M. *et al*. Epitaxial growth of highly compressively strained GeSn alloys up to 12.5% Sn. *J. Cryst. Growth* **384**, 71–76 (2013).

21. Margetis, J. *et al*. Study of low-defect and strain-relaxed GeSn growth via reduced pressure CVD in $H_2$ and $N_2$ carrier gas. *J. Cryst. Growth* **463**, 128–133 (2017).

22. Ghetmiri, S. *et al*. Direct-bandgap GeSn grown on silicon with 2230 nm photoluminescence. *Appl. Phys. Lett.* **105**, 151109 (2014).

23. Wirths, S. *et al*. Lasing in direct-bandgap GeSn alloy grown on Si. *Nature Photon.* **9**, 88–92 (2015).

24. Al-Kabi, S. *et al*. An optically pumped 2.5 μm GeSn laser on Si operating at 110 K. *Appl. Phys. Lett.* **109**, 171105 (2016).

25. Stange, D. *et al*. Optically Pumped GeSn Microdisk Lasers on Si. *ACS Photonics* **3**, 1279–1285 (2016).

26. Al-Kabi, S. *et al*. Optical characterization of Si-based $Ge_{1-x}Sn_x$ alloys with Sn compositions up to 12%. *J. Electron Mater.* **45**, 2133–2141 (2016).

27. Margetis, J. *et al*. Growth and Characterization of Epitaxial $Ge_{1-x}Sn_x$ Alloys and Heterostructures Using a Commercial CVD System. *ECS Trans.* **64**, 711–720 (2014).

28. Gencarelli, F. *et al*. Low-temperature Ge and GeSn Chemical Vapor Deposition using $Ge_2H_6$. *Thin Solid Films* **520**, 3211–3215 (2012).

29. Chen, R. *et al*. Demonstration of a Ge/GeSn/Ge quantum-well microdisk resonator on silicon: enabling high-quality Ge(Sn) materials for micro-and nanophotonics. *Nano Lett.* **14**, 37–43 (2014).

30. Von Den Driesch, N. *et al*. SiGeSn Ternaries for Efficient Group IV Heterostructure Light Emitters. *Small* 1603321 (2017).

31. F. Kroupa. Dislocation Dipoles and Dislocation Loops. *Journal de Physique Colloques* **27**, C3-154–C3-167 (1966).

32. Pezzoli, F. *et al*. Disentangling nonradiative recombination processes in Ge micro-crystals on Si substrates. *Appl. Phys. Lett.* **108**, 262103 (2016).

33. Casey Jr. H. C. Temperature dependence of the threshold current density in InP-$Ga_{0.28}In_{0.72}As_{0.6}P_{0.4}$ ($\lambda$=1.3 μm) double heterostructure lasers. *J. Appl. Phys.* **56**, 1959–1964 (1984).

34. Miller, R. C. Laser oscillation with optically pumped very thin $GaAs$-$Al_xGa_{1-x}As$ multilayer structures and conventional double heterostructures. *J. Appl. Phys.* **47**, 4509–4517 (1976).

35. Sun, G., Soref, R., & Cheng, H. H. Design of an electrically pumped SiGeSn/GeSn/SiGeSn double-heterostructure mid infrared laser. *J. Appl. Phys.* **108**, 033107 (2010).

36. Sun, G., Soref, R., & Cheng, H. H. Design of a Si-based lattice-matched room-temperature GeSn/GeSiSn multi-quantum-well mid-infrared laser diode. *Opt. Express* **18**, 19957–19965 (2010).

37. Ghetmiri, S. *et al*. Study of SiGeSn/GeSn/SiGeSn structure towards direct bandgap type-I quantum well for all group-IV-optoelectronics. *Opt. Lett.* **42**, 387–390 (2017).

# SUPPLEMENTARY INFORMATION

Table of Contents:

1. Material characterization
2. Band diagram calculation
3. Bandgap characteristic analysis
4. Surface profile characterization
5. Lasing mode pattern calculation
6. General characterization of GeSn laser grown on Si
7. References

1. Material Characterization

The GeSn peak fitting process is as follows: the number of peaks is determined by the outline of the curve, i.e., the peak-feature and the shoulder-feature were counted. Then the Gaussian distribution was used to fit each peak. For sample D, the 2θ-ω scan curve was fitted with two Gaussian peaks of the 1st GeSn layer (blue dash curve) and the 2nd GeSn layer (red dash curve) wherever possible, as plotted in Fig. S1(a); while for sample G, three Gaussian peaks corresponding to the 1st, 2nd, and 3rd GeSn layers were fitted using the same process, as shown in Fig. S1(b). The number of layers was also confirmed by the TEM image. Compared to our previous study on GeSn thin film with compressive strain[1], the measured GeSn peaks in this work shifted towards lower angle due to relaxation of the material (having the same Sn composition). Moreover, the narrower peak line-width of the top GeSn layer compared to that of the bottom GeSn layer(s) indicates its higher material quality, which agrees well with the results from measured threading dislocation density (TDD).

The in-plane ($a_\parallel$) and out-of-plane ($a_\perp$) lattice constants of the GeSn alloys were obtained from XRD-RSM (from (-2 -2 4) plane reflection), as shown in Fig. S2. The dashed red line represents the strain relaxation. By fitting the broadened contour plots, the lattice constants and strain for each GeSn layer were extracted. Note that for sample B, another Ge layer was shown with larger in-plane and smaller out-of-plane lattice constants. This is due to the 10 nm-thick Ge cap layer initially designed for passivation which lately showed negligible effects for the overall device performance.

The Sn compositions were determined by secondary ion mass spectrometry (SIMS) measurements for all samples except for samples A and B as they feature lower Sn composition, therefore their Sn compositions were determined by the XRD measurements with well accepted bowing parameter[2].

The layer thicknesses were determined by TEM images, from which each GeSn layer can be clearly resolved so that the layer thickness can be directly measured. The measured thicknesses were cross checked by SIMS. Figure S3 shows the typical TEM images of samples D and G.

## 2. Band diagram calculation

Based on the obtained Sn composition, strain and layer thickness, the band diagram at 300 K was calculated using the effective mass and matrix propagation method[3,4]. The strain-induced bandgap change[5] were considered for the calculations. The band offsets were obtained based on well acknowledged theoretical calculation[6]. Figure S4 shows the schematic band structures of the samples D and G (not to scale). The $E_{cL}$, $E_{c\Gamma}$, $E_{vhh}$, and $E_{vlh}$ represent energy levels of L- and $\Gamma$-valleys at conduction band (CB), and heavy hole (hh) and light hole (lh) at valence band (VB), respectively. For sample D, the 1$^{st}$ GeSn layer is indirect bandgap due to residual compressive strain while the 2$^{nd}$ GeSn layer is direct bandgap. Since the Sn composition of the 1$^{st}$ GeSn layer is lower relative to the 2$^{nd}$ GeSn layer, the formation of energy barriers at both CB and VB leads to photo-generated carriers confined in the 2$^{nd}$ GeSn layer where the band-to-band recombination (either PL or lasing) occurs. The direct bandgap energy of 0.476 eV agrees well with the measured PL peak position of 2610 nm. For sample G, all three GeSn layers feature direct bandgap, with $E_{g3} < E_{g2} < E_{g1}$ ($E_{gn}$ is bandgap energy of n$^{th}$ GeSn layer). The observed PL peak at 3225 nm is consistent with the calculated bandgap energy of the 3$^{rd}$ GeSn layer at 0.385 eV, indicating the excellent carrier confinement in the 3$^{rd}$ GeSn layer. Based on our calculation results, all samples feature type-I band alignment with the carriers confined in the top GeSn layer. The PL measurements confirmed that the observed emission wavelengths match the bandgap energies of the top GeSn layers. The bandgap energy calculations are summarized in Table S1.

## 3. Bandgap characteristic analysis

The temperature-dependent PL measurements were performed on each sample. We analyzed the bandgap characteristic based on a systematic study of the PL spectra and then draw the conclusion of the material bandgap directness. Figure S5 shows the temperature-dependent normalized integrated PL intensity of each sample. For sample A, as temperature decreases from 300 to 150 K, the integrated PL intensity decreases due to the reduced number of carriers populating on $\Gamma$ valley. Below 150 K the decreased non-radiative recombination velocity overcompensates the reduced number of thermal activated carriers, resulting in the increased integrated PL intensity at temperatures from 150 to 10 K. Since the bandgap theoretical calculation indicates that there is almost no energy difference between the $\Gamma$ and L valleys, therefore, the bandgap of sample A is at the indirect-to-direct transition point. For samples B, C, and D, the bandgap theoretical calculation indicates that their $\Gamma$ valleys are lower than L valleys. As the temperature decreases, the integrated PL intensity monotonically increases. The increases of 6-, 2-, and 4-times were observed for samples B, C, and D, respectively. For samples E, F, and G, whose Sn compositions are far beyond the theoretical predicted indirect-to-direct transition point, their integrated PL intensity significantly increased at lower temperature. The increases of 65-, 24-, and 25-times were observed for samples E, F, and G, respectively, which is attributed to the further lowered $\Gamma$ valley for dominant direct bandgap transition. The bandgap energy separation between $\Gamma$ and L valleys increases as the Sn composition increases, which is a typical direct bandgap material behavior.

Note that the 65-times PL intensity increase at low temperature was observed for sample E, which is the highest value among all samples. However, based on material characterization, sample E features relative lower material quality. This can be explained as following: due to the lower material quality of sample E, the PL intensity at room temperature is weak, resulting in a larger ratio of low and high temperature PL intensities.

## 4. Surface profile characterization

To investigate the surface roughness, the atomic-force microscopy (AFM) characterizations have been performed on as-grown samples and post-etch samples. Note that for the measurements of post-etch samples, the scan area was on the etched surface not on sidewall since it is very difficult to scan the sloped surface. The validity of this measurement method was indicated by SEM image (Fig. S6) which showed that their surface profiles are very close. The results are summarized in Table S2. (Note, the etched sample G was broken during the measurement therefore the data currently is not available. However, based on the available data, a clear conclusion could be drawn.)

Based on the obtained data, the surface roughness is partially due to the initial material growth and strain release. For samples A, B, and E, the wet etching also adds additional roughness. While for other samples, either roughness does not change or even reduces after etching. According to the reported data of surface roughness[7], the measured surface roughness in this work is with comparable value.

We have estimated surface roughness related loss to be ~10 dB/cm based on the scattering loss study for Si waveguide operating at 1.55 µm with similar roughness[8]. Since our GeSn lasers operate at longer wavelength, the actual waveguide loss could be lower than this value. The loss is at reasonable value for early F-P cavity lasers.

## 5. Lasing mode pattern calculation

Figure S6(a) shows the pattern of the fundamental transverse electric (TE$_0$) mode obtained using a 2D mode solver[9]. The cross-section of the layout was consistent with the fabricated device, i.e., the top and bottom width of the ridge waveguide were 2 and 5 μm, respectively, and the etch depth was 800 nm. The calculations were done for a layer structure corresponding to sample D. The refractive index ($n$) of GeSn was obtained from ref. 10. Since the $n$ of Ge$_{0.9}$Sn$_{0.1}$ is very close to that of Ge$_{0.87}$Sn$_{0.13}$, an identical $n$ of 4.25 was used for the entire GeSn layer. Mode overlap with the GeSn and the Ge layer of 85.2% and 14.4% respectively were obtained, revealing the superior optical confinement in the GeSn layer. In comparison, the waveguide with the same layout except the 90° sidewall was simulated as well. Under the same configuration of $n$, a slightly increased mode overlap with GeSn of 85.9% was obtained, as shown in Fig. S6(b). This indicates that as long as the smooth surface was achieved, the deterioration of device performance due to sloped sidewall can be ignored in this work.

## 6. General characterization of GeSn laser grown on Si

The laser-output versus pumping-laser-input (L-L) curves for all lasers operating at 77 K are shown in Fig. S7(a). The clear threshold feature was observed for each curve, confirming the lasing characteristic.

Figures S7(b1)-(b4) show the characteristic temperatures (T$_0$) of samples B, C, D, and F. The T$_0$ was obtained by the linear fitting using equation ln ($I_{th}$) ∝ (T/T$_0$), where T and $I_{th}$ are temperature and corresponding threshold, respectively.

The major efforts towards further improving the GeSn double heterostructure (DHS) laser performance in terms of higher operating temperature and lower threshold will be the optimization of the ridge waveguide structure to achieve single-mode operation, and the fabrication of a thicker, defect-free GeSn top layer to maximize the mode overlap with it. Moreover, the material growth study revealed that there is still room to further improve the material quality by optimizing the growth conditions.

# 7. References

1. Ghetmiri, S. *et al*. Direct-bandgap GeSn grown on silicon with 2230 nm photoluminescence. *Appl. Phys. Lett.* **105**, 151109 (2014).

2. Ryu, M. *et al*. Temperature-dependent photoluminescence of Ge/Si and $Ge_{1-y}Sn_y$/Si, indicating possible indirect-to-direct bandgap transition at lower Sn content. *Appl. Phys. Lett.* **102**, 171908 (2013).

3. Chang, G. *et al*. Strain-Balanced $Ge_zSn_{1-z}-Si_xGe_ySn_{1-x-y}$ Multiple-Quantum-Well Lasers. *IEEE J. Quant. Electron.* **46**, 1813 (2010).

4. Jaros, M. Simple analytic model for heterojunction band offsets. *Phys. Rev. B, Condens. Matter.* **37**, 7112 (1988).

5. Menéndez, J. *et al*. Type-I $Ge/Ge_{1-x-y}Si_xSn_yGe$ / $Ge_{1-x-y}Si_xSn_y$ strained-layer heterostructures with a direct GeGe bandgap. *Appl. Phys. Lett.* **85**, 1175 (2004).

6. Sun, G., Soref, R., & Cheng, H. H. Design of an electrically pumped SiGeSn/GeSn/SiGeSn double-heterostructure mid infrared laser. *J. Appl. Phys.* **108**, 033107 (2010).

7. Wirths, S. et al. Reduced Pressure CVD Growth of Ge and $Ge_{1-x}Sn_x$ Alloys. *ECS J. Solid State Sci. Technol.*, 2 N99 (2013).

8. Lee, K. *et al*. Fabrication of ultralow-loss Si/SiO$_2$ waveguides by roughness reduction. *Opt. Lett.* **26**, 1888 (2001).

9. Fallahkhair, A. *et al*. Vector Finite Difference Modesolver for Anisotropic Dielectric Waveguides. *J. Lightwave Technol.* **26**, 1423 (2008).

10. Tran, H. *et al*. Systematic study of $Ge_{1-x}Sn_x$ absorption coefficient and refractive index for the device applications of Si-based optoelectronics. *J. Appl. Phys.* **119**, 103106 (2016).

Figures

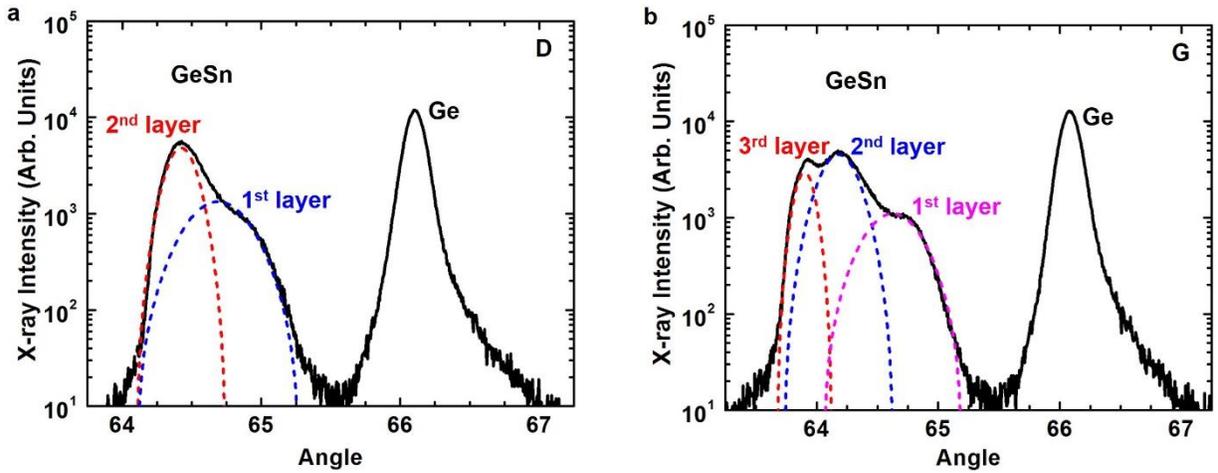

Fig. S1 Fitting process of 2θ-ω scan for samples (a) D and (b) G. The multiple peaks are clearly resolved, indicating the existence of multiple GeSn layers with different Sn compositions.

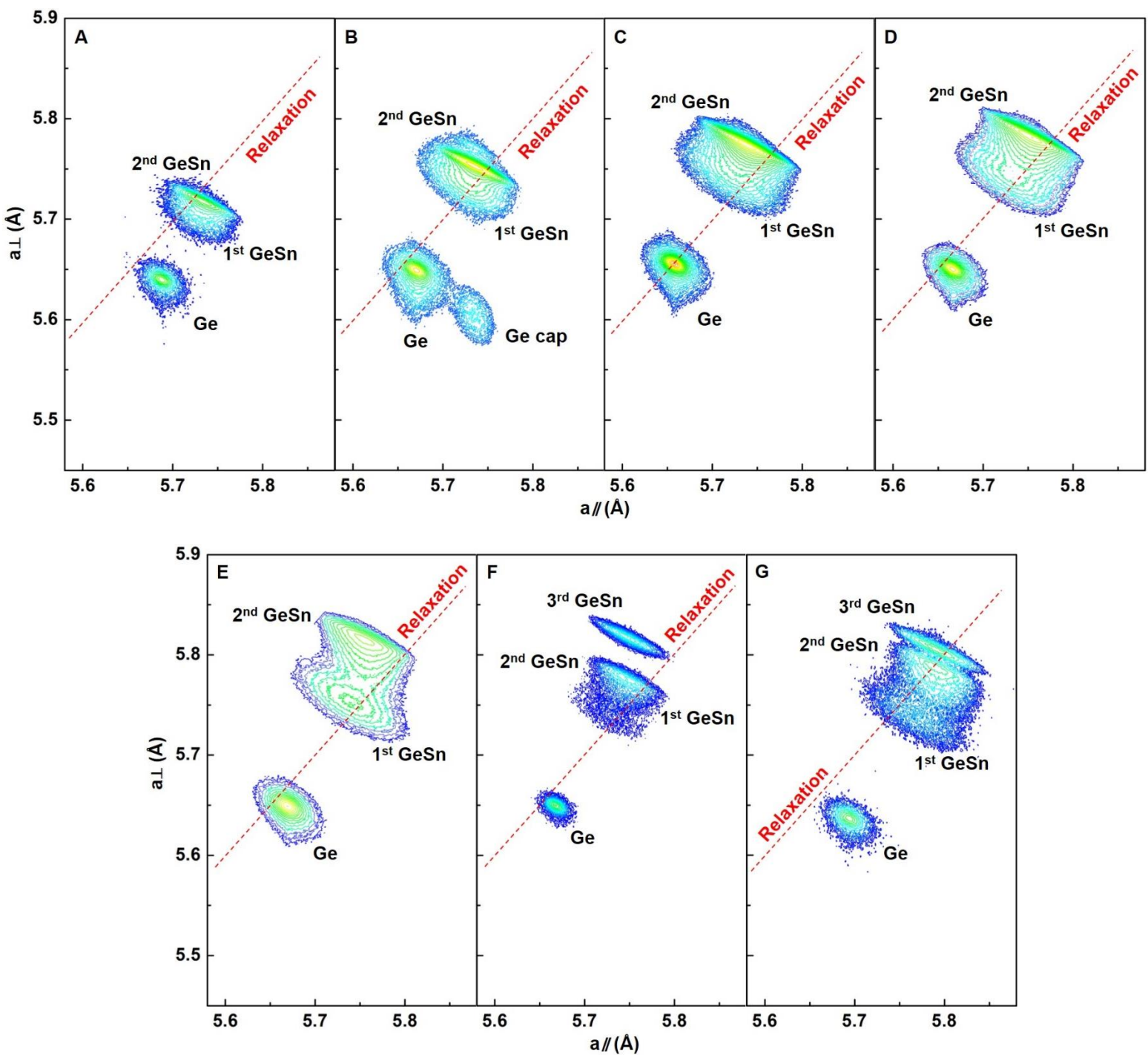

Fig. S2 XRD-RSM of each sample shows the in-plane (a∥) and out-of-plane (a⊥) lattice constants as well as the strain. The broadened contour plot of GeSn indicates the multi-layer feature of the GeSn.

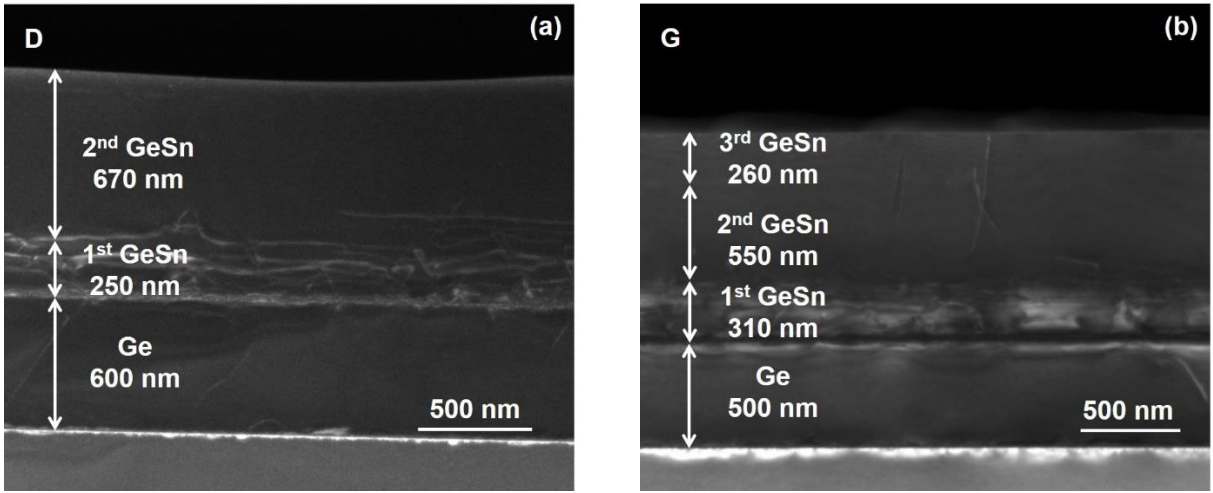

Fig. S3 TEM images of samples (a) D and (b) G. Each layer can be clearly resolved. The top GeSn layers feature low-density TD.

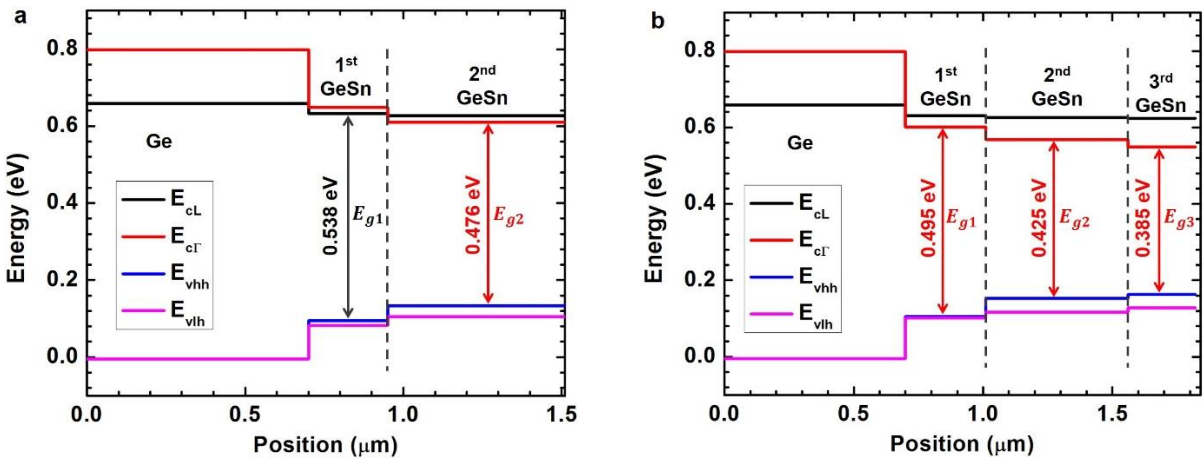

Fig. S4 Band diagram calculation for samples (a) D and (b) G at 300 K (not to scale).

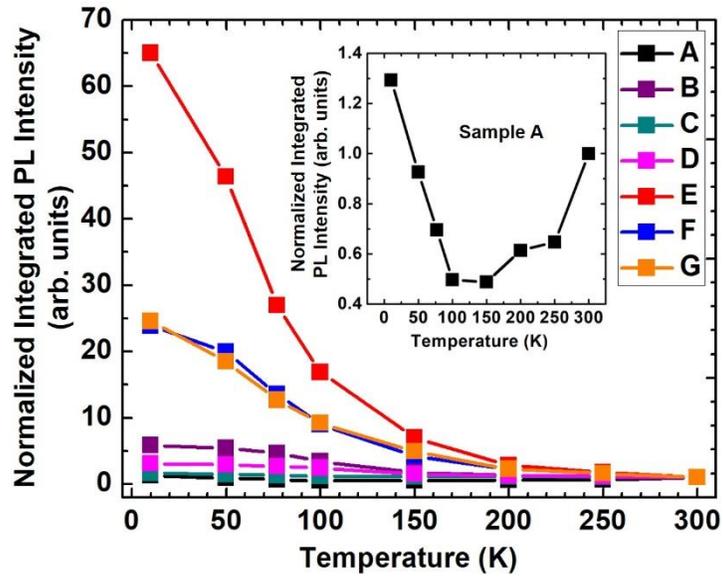

Fig. S5 Temperature-dependent normalized integrated PL intensity of each sample showing the typical direct bandgap material characteristics. Inset: Details of sample A.

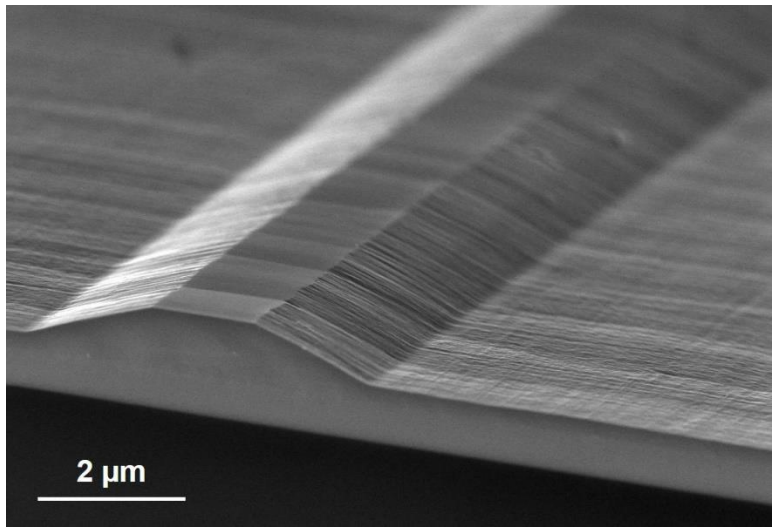

Fig. S6 SEM images of the laser device showing the sidewall of the waveguide.

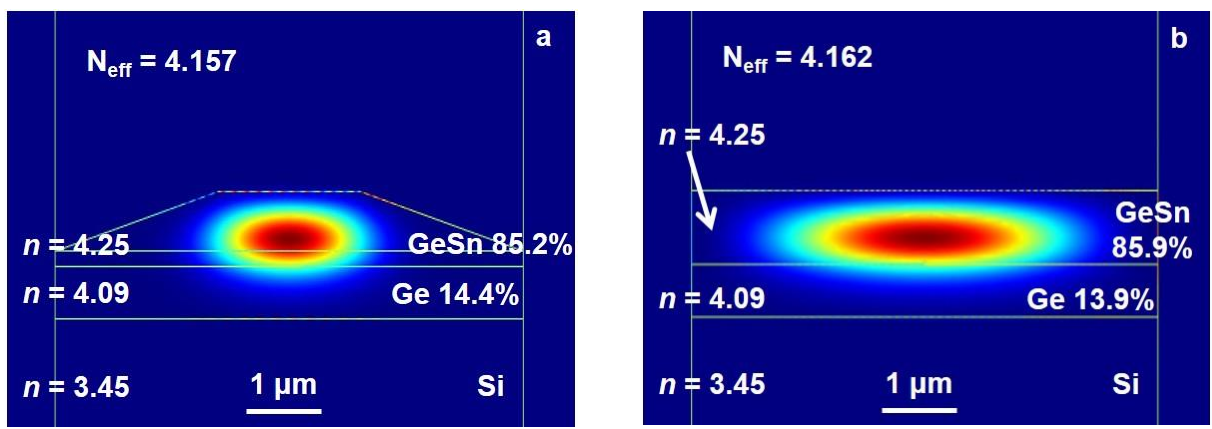

Fig. S7 Calculated pattern of the fundamental transverse electric mode for (a) actual device and (b) ideal ridge waveguide with 90º sidewall. The $N_{eff}$ is the effective index. The mode overlap difference with the GeSn layer is only 0.7% between the two structures.

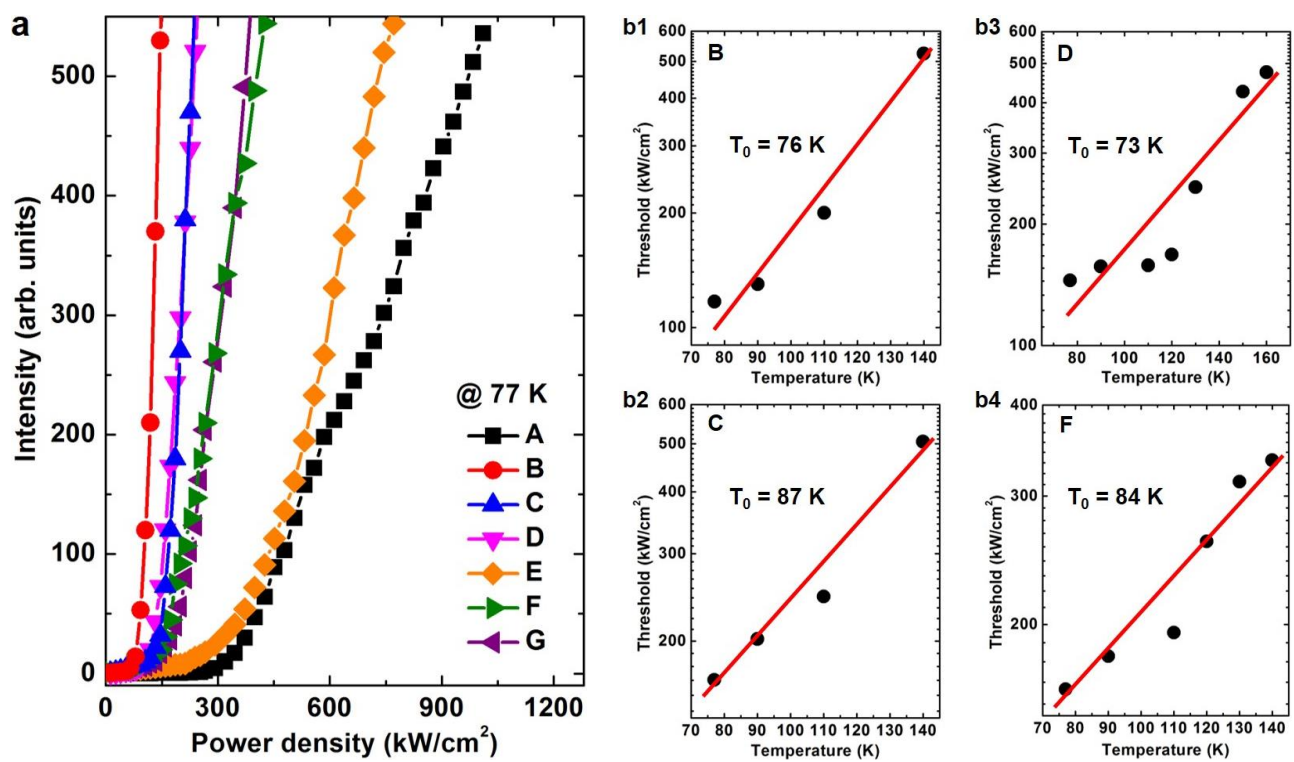

Fig. S8 (a) L-L curves of GeSn lasers at 77 K. (b1)-(b4) Fitted $T_0$ of GeSn lasers.

Tables

Table S1. Summary of bandgap energy for each GeSn layer

|  | A | B | C | D | E | F | G |
|---|---|---|---|---|---|---|---|
| $E_{g1}$ (eV) | 0.595 | 0.537 | 0.536 | 0.538 | 0.464 | 0.541 | 0.495 |
| $E_{g2}$ (eV) | 0.566 | 0.490 | 0.489 | 0.476 | 0.390 | 0.484 | 0.425 |
| $E_{g3}$ (eV) |  |  |  |  |  | 0.408 | 0.385 |

Table S2 Summary of surface roughness (unit: nm)

| Sample | A | B | C | D | E | F | G |
|---|---|---|---|---|---|---|---|
| As-grown roughness | 3.75 | 3.94 | 9.12 | 10.30 | 5.08 | 18.00 | 9.34 |
| Post-etch roughness | 6.27 | 10.40 | 7.17 | 10.60 | 13.70 | 6.03 | N.A. |